\documentclass[11pt]{article}
\usepackage{epsfig}
\usepackage{wrapfig}
\usepackage{subfigure}

\begin{document}

\title{Estimation of Weighting Potential for a TPC}

\author{S. Mukhopadhyay$^a$, N.Majumdar$^a$, R. Veenhof$^b$}
\date{\small{$^a$ INO Section, SINP, Kolkata, India\\
$^b$ CERN CH-1211, Geneve, Switzerland\\
supratik.mukhopadhyay@saha.ac.in, nayana.majumdar@saha.ac.in,\\
Rob.Veenhof@cern.ch}}
\maketitle

\begin{abstract}
In this work, we have computed the three dimensional weighting potential, field
and pad response function (PRF) of typical time projection chambers (TPC) using
a recently formulated and developed boundary element method solver,
namely, the nearly exact BEM solver. A realistic geometry of the device is
found to have significant influence on the estimation of signal generation.
\end{abstract}

\section{Introduction}
Over the last three decades, the time projection chamber (TPC)
\cite{Ambrosio2005} has been considered as one of the
important and successful gas detectors, especially in accelerator based
experiments. The signal in a TPC can be
estimated using the Shockley-Ramo \cite{Shockley,Ramo} theorem following which
the electric and weighting fields turn out to be two fundamental quantities.

In this work, we have computed the weighting potential of a typical TPC using
a recently developed boundary element method (BEM) solver, namely, the nearly
exact BEM (neBEM) solver \cite{EABE2006,NIMA2006}. Weighting potential, field
and pad response function of TPC-s of realistic geometries have been accurately
estimated using this solver. Results for a two-dimensional strip detector for
which an analytic solution exists (thereby neglecting the presence of the anode
wires altogether) or those obtained by using elegantly derived semi-empirical
expressions \cite{Gatti79}, cannot be considered to be accurate
for all possible geometries. As is evident from our results, presence of the
anode wires and three-dimensionality of the detectors do alter the weighting
potential considerably.

In this regard, the finite element method (FEM) packages (commercial or
otherwise) are known to perform poorly in spite of consuming large amount of
computational resources. On the other hand, conventional BEM solvers are also
known to
suffer from several drawbacks. In the present formulation, many of the drawbacks
of BEM have been removed and, as a result, the neBEM solver can be used very
effectively to solve for problems related to estimating the electrostatic
configuration of TPCs.

\section{Background}
According to the Shockley-Ramo theorem, the current induced at time $t$ on an
electrode due to a charge at position $\vec r$ can be evaluated as follows:
\begin{equation}
i(t) = q \vec{v} \vec{E_w}
\end{equation}
where $q$ represents the charge, $\vec v$ is its velocity and $\vec{E_w}$ is the
weighting field at $\vec r$ associated with the electrode under study.
Similarly, PRF is also used for estimating signal induced on cathode pads.
It is
necessary to be able to compute the mentioned parameters to an acceptable
accuracy in 3D. It may be mentioned here that the final aspects of signal
generation can be very effectively modeled by Garfield \cite{Garfield}.

\section{Results and discussions}
\subsection{Comparison with analytical solutions}
A simple closed-form expression for the weighting potential exists for a 2D
strip detector with no gap between strips. It can be written as \cite{Rehak99}
\begin{equation}
\label{eq:AnalyticPotential}
\Phi(x,y) \,=\, \frac{1}{\pi}
	( \arctan( \tanh(\beta) \cdot \tan(\gamma) )
		- \arctan( \tanh(\alpha) \cdot \tan(\gamma) ) )
\end{equation}
where
\[
\alpha = \pi \frac{x - a/2}{2d};\, \beta = \pi \frac{x+a/2}{2d};\,
\gamma = \pi \frac{y}{2d}
\]
The surface plots of weighting potentials obtained using the neBEM solver have
been presented in fig.\ref{fig:StripPotNumeric}. Qualitatively, the comparison
with the analytic result \cite{Rehak99} is found to be very satisfactory.

\subsection{Effect of the presence of anode wires}
In order to estimate the effect of anode wires on the electrostatic
configuration, we have included a plane of wires within the simple geometry
considered above (termed \textit{wired}).

\begin{figure}[h]
\centering
\subfigure[Weighting potential]{\label{fig:StripPotNumeric}\includegraphics[width=0.45\textwidth]{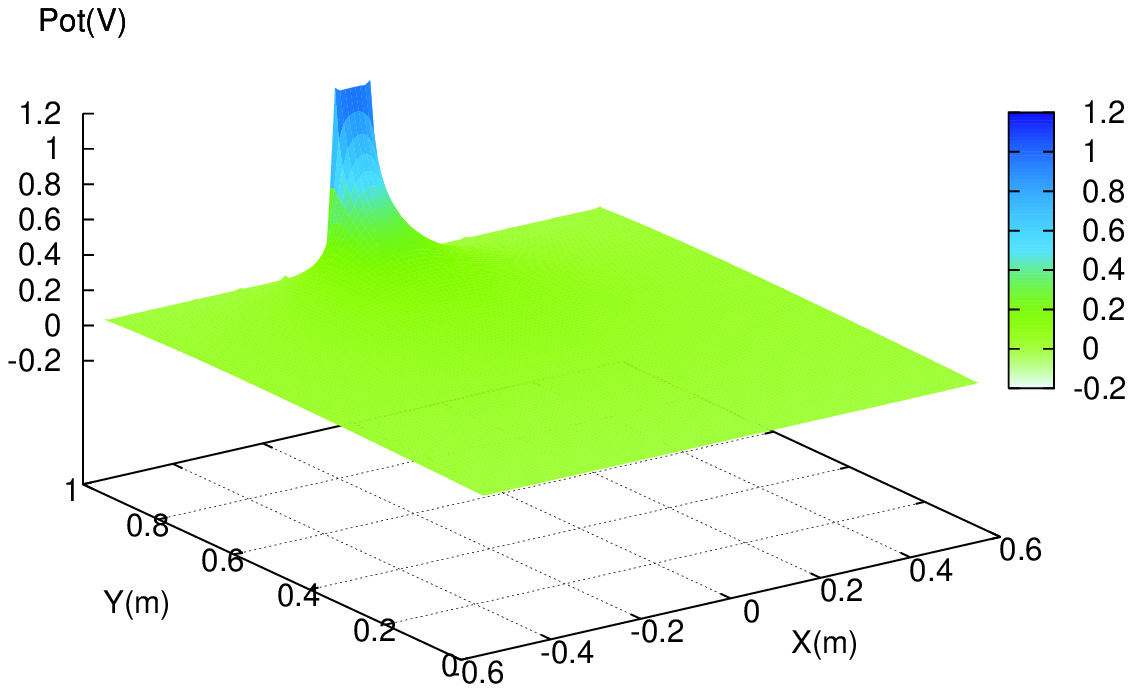}}
\subfigure[Potential distribution]{\label{fig:AnalyticAnodeWirePotVsY}\includegraphics[width=0.45\textwidth]{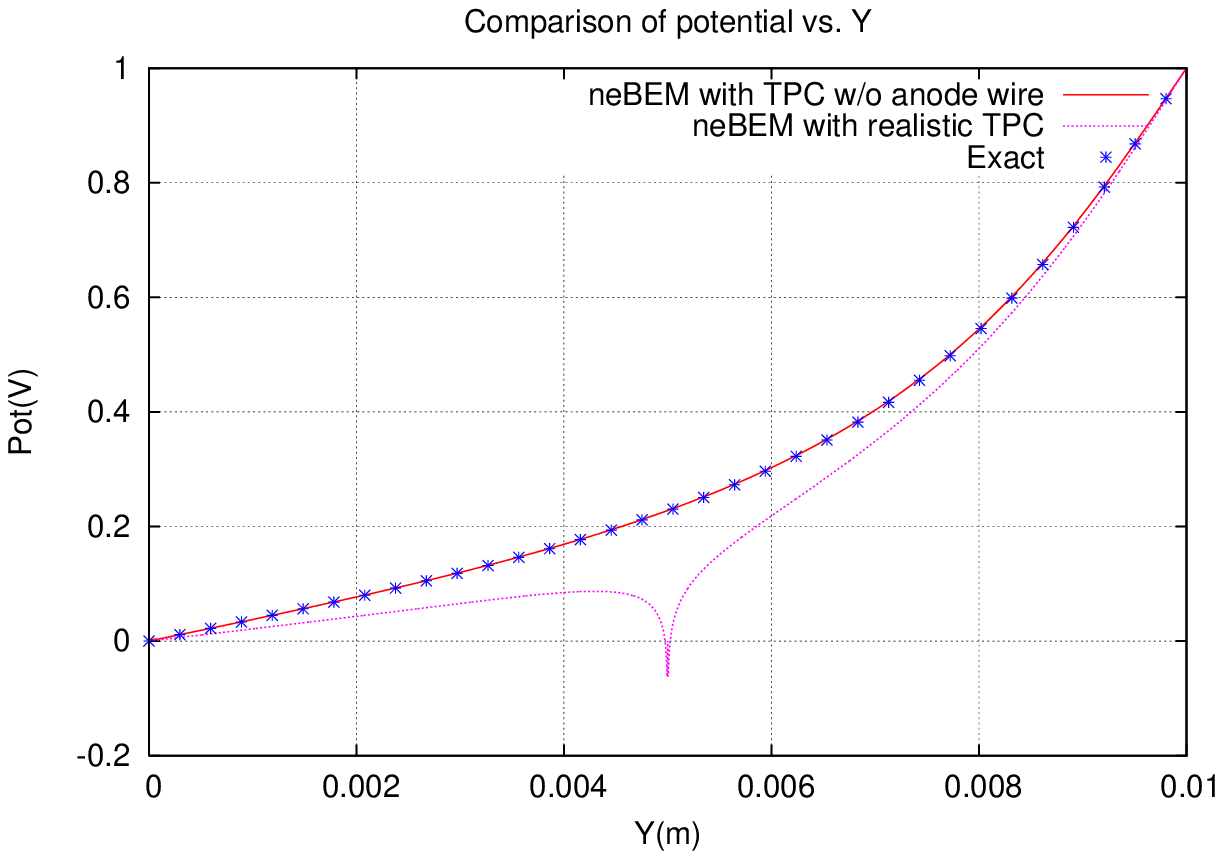}}
\caption{(a) Weighting potential surface using neBEM and
(b) comparison of potential distribution of a simplified strip detector
computed using neBEM with analytic solution \cite{Rehak99}}
\label{fig:Potentials}
\end{figure}

The weighting potential along Y for the \textit{unwired} simple detector
naturally
matches perfectly with the exact solution. The \textit{wired} detector,
however, has a significantly different weighting potential distribution
throughout, the difference being very large near the anode wire
(fig.\ref{fig:AnalyticAnodeWirePotVsY}). This is logical because the
anode wire provides a new zero potential surface. This deviation creates a
large electric field, Ey, near the anode wire which, naturally, influences the
generated signal.
Thus, it is noted,that the naive use of analytic expressions
may lead to non-negligible errors.

\subsection{Computations for a realistic TPC}
\begin{figure}[h]
\centering
\subfigure[Pad response function]{\label{fig:PRFCompare}\includegraphics[width=0.45\textwidth]{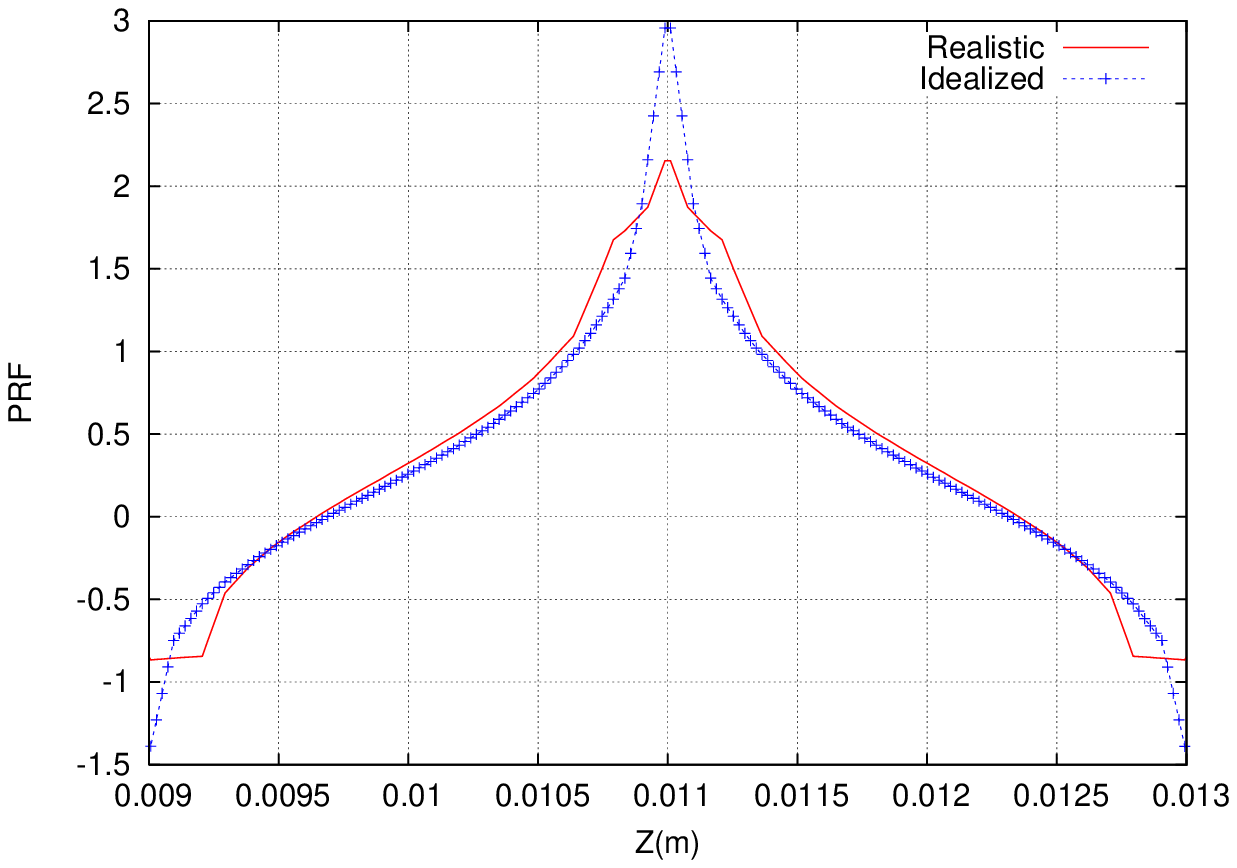}}
\subfigure[Potential distribution]{\label{fig:ThreeDEffectWtPot}\includegraphics[width=0.45\textwidth]{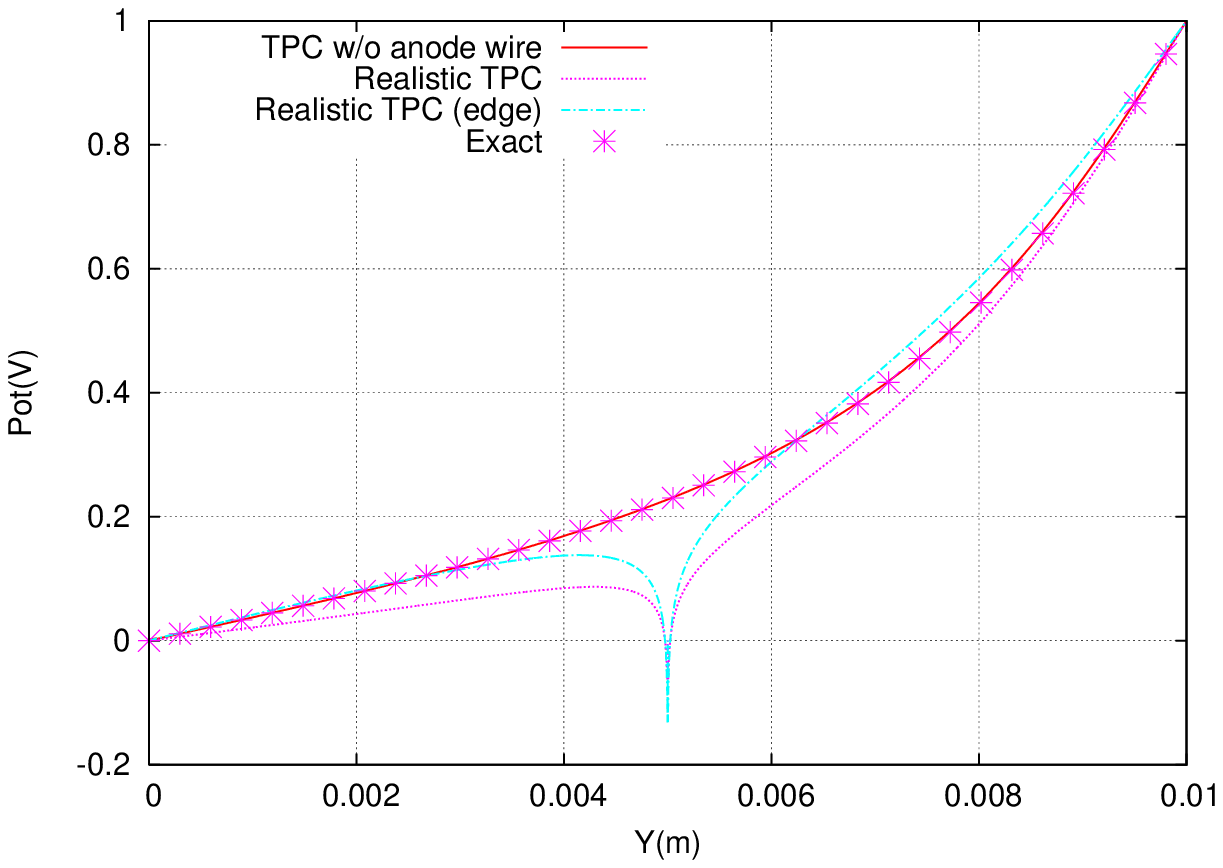}}
\caption{(a) One dimensional PRF distributions for the gap-less and the
realistic TPCs and (b) Comparison of weighting potentials for 2D strip detector,
mid-plane of a realistic TPC and edge of the same TPC}
\label{fig:PRFand3D}
\end{figure}

Next, we have considered a \textit{realistic} TPC where the surface
representation of the
cathode wire plane has been corrected. We have also considered gaps between
segments in this geometry. In the \textit{idealized} TPC, we have ignored these
gaps. As preliminary estimation, we have computed the one-dimensional PRF using
the weighting field distribution and presented the results for the two TPCs in
fig.\ref{fig:PRFCompare}. It can be seen that omission of gaps between
segments of the cathode plane can lead to small, but finite, error in estimating
the PRF as well.

\subsection{Three dimensional effects}
In fig.\ref{fig:ThreeDEffectWtPot}, we show a comparison between the weighting
potentials of a 2D strip detector (analytic solution), the same at the mid-plane
of a realistic (anode wires added to the analytic geometry) TPC detector, in
addition to the values obtained at the edges of the same realistic TPC detector.
The significant amount of difference between the
mid-plane and the edge values is immediately apparent. This difference
emphasizes even more the importance of precise computation of electrostatic
configuration for a TPC.

\section{Conclusion}
Accurate 3D weighting potential and fields have been calculated using the
recently developed neBEM solver. Now,
it should be easy for us to simulate the charge / signal induced on any
electrode of a detector due to the passage of an ionizing particle passing
through a detector by calculating the real field within the detector (can be
done using the present solver) and the drift velocity (can be done using
\cite{Garfield}).

\end{document}